\documentclass{article}

\usepackage{arxiv}

\usepackage[utf8]{inputenc} % allow utf-8 input
\usepackage[T1]{fontenc}    % use 8-bit T1 fonts

\usepackage{hyperref}
\usepackage{graphicx}
\usepackage{longtable}
\usepackage{multirow}
\usepackage{array}
\usepackage{enumerate}
\usepackage{caption}
\usepackage{paralist}            
\usepackage{enumitem}
\usepackage{verbatim}

\title{A Holistic View on Data Protection for Sharing, Communicating, and Computing Environments: Taxonomy and Future Directions}

\author{ \href{https://orcid.org/0000-0003-3746-6034}{\includegraphics[scale=0.06]{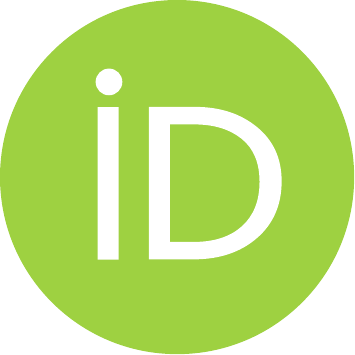}\hspace{1mm}Ishu Gupta*}
	\\
	Cloud Computing Research Center\\
	Department of Computer Science and Engineering\\ 
	National Sun Yat-sen University\\
	Kaohsiung, Taiwan\\
	\texttt{ishugupta23@gmail.com} \\
	\And
	\href{https://orcid.org/0000-0002-8053-5050}{\includegraphics[scale=0.06]{orcid.pdf}\hspace{1mm}Ashutosh Kumar Singh} \\
	Department of Computer Applications\\
	National Institute of Technology\\
	Kurukshetra, India \\
	136119\\
	\texttt{ashutosh@nitkkr.ac.in} \\
}

\renewcommand{\shorttitle}{\textit{Data Protection for Sharing, Communicating, and Computing Environments}}

\hypersetup{
	pdfsubject={q-bio.NC, q-bio.QM},
	pdfauthor={David S.~Hippocampus, Elias D.~Striatum},
	pdfkeywords={First keyword, Second keyword, More},
}

\begin{document}
	\maketitle
	
	\begin{abstract}
		The data is an important asset of an organization and it is essential to keep this asset secure. It requires security in whatever state is it i.e. data at rest, data in use, and data in transit. There is a need to pay more attention to it when the third party is included i.e. when the data is stored in the cloud then it requires more security. Since confidential data can reside on a variety of computing devices (physical servers, virtual servers, databases, file servers, PCs, point-of-sale devices, flash drives, and mobile devices) and move through a variety of network access points (wireline, wireless, VPNs, etc.), there is a need of solutions or mechanism that can tackle the problem of data loss, data recovery and data leaks. In this context, the paper presents a holistic view of data protection for sharing and communicating environments for any type of organization. A taxonomy of data leakage protection systems and major challenges faced while protecting confidential data are discussed. Data protection solutions, Data Leakage Protection System's analysis techniques, and, a thorough analysis of existing state-of-the-art contributions empowering machine learning-based approaches are entailed. Finally, the paper explores and concludes various critical emerging challenges and future research directions concerning data protection.
	\end{abstract}

	% keywords can be removed

	\section{Introduction}
	Data is recognized as the most vital asset of an enterprise because it defines each organization’s uniqueness \cite{JISE}. In any organization, data is the main foundation of information, knowledge and ultimately the wisdom for correct decisions and actions \cite{OnILIS,Nishad}. It might be helping to cure a disease, boost a company’s revenue, make a building more efficient or be responsible for those targeted as we keep seeing \cite{GUIM-SMD}. If the data is relevant, complete, accurate, timely, consistent, meaningful, and usable, then it will surely help in the growth of the organization \cite{Batra,DT-ILIS}. Companies that do not understand the importance of data management are less likely to survive in the modern economy. Therefore, it is essential to understand the importance of data management in companies \cite{RL2021,Madan}. Data should be carefully cultivated, managed, and refined into information that will allow an enterprise to better serve their community and ensure to remain viable in today’s competitive landscape \cite{AA2013,Vartika}.
	
	In the real world scenario, a data distributor has to share the sensitive data of an entity among various stakeholders such as business partners, customers, and employees within or outside the organization's premises for doing business \cite{MLPAM,Ayushi}. But the receiving entity may misuse this data and can leak it deliberately or inadvertently to some unauthorized third party \cite{IJAST,Jadon}. Data leakage is the intentional or unintentional distribution of confidential or private information or data to an unauthorized malicious entity \cite{Godha,Ankit}. Critical data in various organizations include patient information, Intellectual Property (IP), personal information like credit card data, financial information and various other information depending upon the organization \cite{Saxena,SELI}.
	
	Data leakage can cause a serious threat to the organization's confidentiality as the number of incidents as well as the cost procured due to these leakages to continue to increase \cite{Tiwari}. It poses a great challenge for any organization \cite{Khushbu,Arora}. It can destruct the enterprise's reputation and goodwill, diminish shareholder's value and decline the firm's status and rank \cite{Sharma,Sloni}. Mostly precarious threats that any company faces, are because of insiders in the company, after all, as the insiders know the company’s internal system.  To handle the attacks caused by insiders of the company is a most difficult task \cite{Kesharwani,Jalwa}.
	
	It is essential to protect the confidential information as it increases the risk of falling the sensitive information in unauthorized hands and then it can be misused by unauthorized third party \cite{PCS,Kaur2018}. Thus, it has become consequential for any organization to detect and prevent such leakage. Conversely, limiting the sharing of data in order to preserve the security and privacy of confidential information might reduce the organization's performance. It influences the potential to perform the operation that can serve best to the organization and its customers \cite{CC}. Data leakage detection and prevention significantly play an inevitable role ineffective protection of the data. In this paper, concepts, challenges, existing solutions, research gaps, and future directions for sharing, communicating, and computing environments for any type of organization are discussed in detail.  
	
	\textit{Organization}: Section 2 discusses the motivation for carrying out this study. Section 3 entails the challenges faced by the data protection system while protecting the data followed by the categorization of data protection solutions in section 4. Section 5 describes the concept of data leakage protection systems (DLPTSs) including data states, deployment schema, and DLPTSs analysis technique. An analysis and comparison of existing approaches and solutions along with their strengths and weaknesses are described comprehensively in section 6. Thereafter, the emerging challenges and future research direction in the field are entailed in section 7 followed by the conclusion and future scope of the work in section 8.
	
	\section{Motivation}
	In recent years, the internet and related technologies have grown rapidly \cite{Animesh,Hybrid}. It offered the unrivaled capability to access and redistribute the digital data \cite{IDS,Harsh}. It is easy to copy a huge amount of digital data at almost no cost that can be transferred in a very short time via the internet \cite{JCOMSS,IJNSA}. The various organizations use this facility to enhance their capability by transferring data from one place to another but it involves a number of threats in transferring the confidential data as it can be made public by any malicious entity \cite{Preetesh}.  
	
	According to an exotic chronology of data breaches explored by Privacy Rights Clearinghouse (PRC), $90,61,64,759$ records have been breached in the united state alone from $5, 278$ data breaches made public since $2005$ \cite{PRC}. It is not hard to speculate that it is just a fingertip of an iceberg as most of the time, the data leakage cases are not recorded due to concern of restrictive penalties and loss of customer trust. In the year $2016$, $4, 004, 004$ records have been breached in the health care, medical providers, and medical insurance services organization alone and $300$ breaches made public \cite{CDB}.
	
	According to the annual cost of data breach study conducted by Ponemon Institute, the average consolidated cost of a data breach is $\$4$ million. This study states that the cost procured for each stolen or lost record consisting of confidential and sensitive information increased from a consolidated average of $\$154$ to $\$158$ \cite{SI}. In highly regulated industries like healthcare, data breach cost is $\$355$ per record which is $\$100$ more than in $2013$ \cite{CNBC}. Because of these reasons; the data leakage problem is increasing day by day. 
	
\begin{figure}[!ht]
		\centering
		\includegraphics[width=\textwidth]{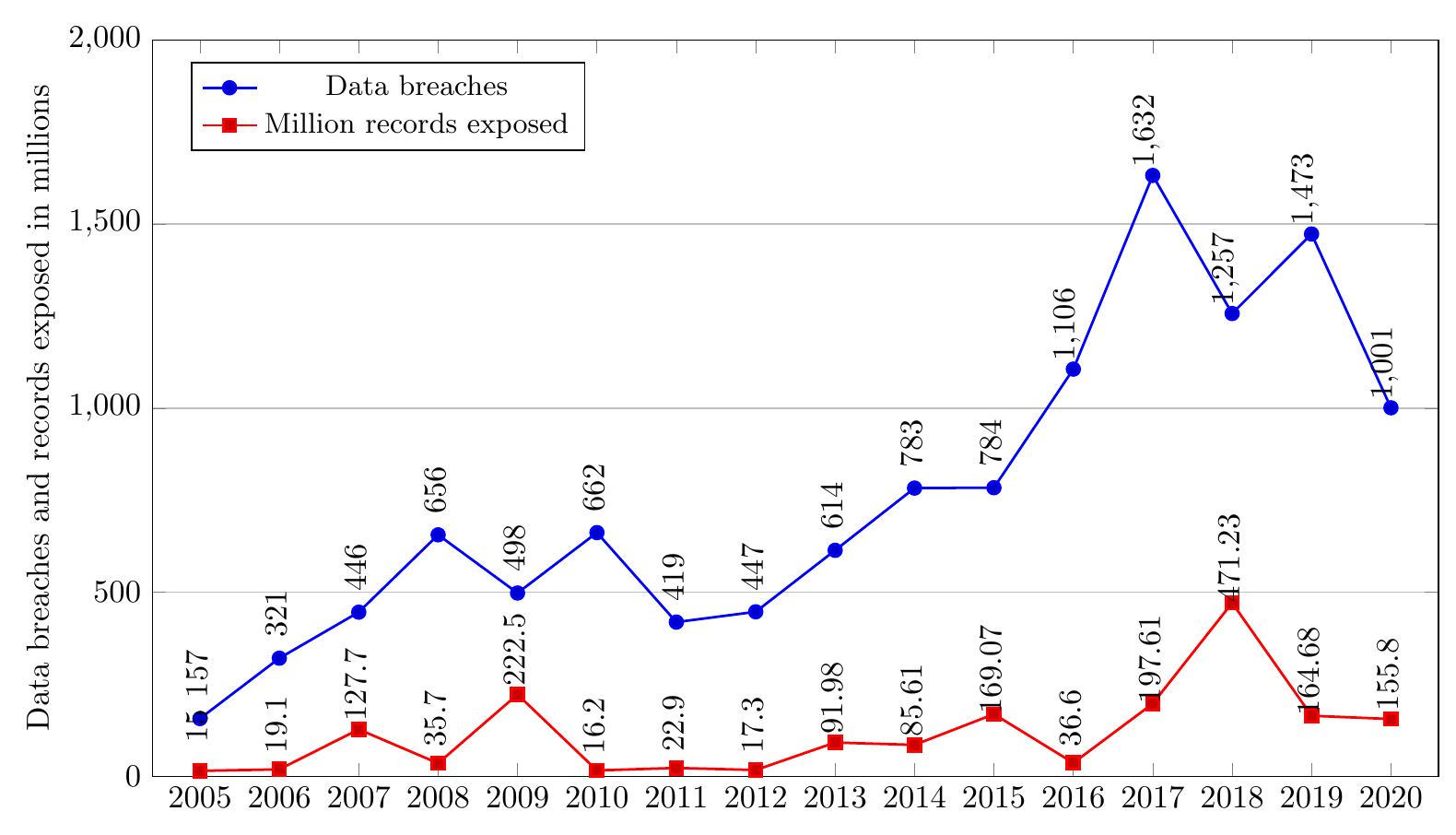}
		\caption{Annual number of data breaches and exposed records in the United States from 2005 to 2020 (in millions).}
		\label{fig2}
	\end{figure}

	A study reports that the number of leaked confidential data records has become $10$ times within $4$ years and it reached up to $1:1$ billion in $2014$ and it is kept on increasing as the number of users are increasing, also the malicious entity \cite{ABER}. Fig. \ref{fig2} \cite{TRC2021} presents the development of cyberattacks over time. It presents the recorded number of data breaches and records exposed in the United States between $2005$ and $2020$. In $2020$, the number of data breaches in the United States amounted to $1001$ with $155.8$ million records exposed which is much higher compared to the past years.\\
	The data leakage problem has reached a new dimension. There is a need for a general mechanism that can handle the data leakage problem by preventing the data leakage and by identifying the malicious entity responsible for data leakage. As the data is shared among multiple entities thus it becomes difficult to identify the malicious entity who has leaked the confidential information \cite{Guilty,Gautam}. In this paper, comprehensive solutions are explored that solve the data leakage problem by preventing the leakage and identifying the malicious entity who has leaked the organizational sensitive data. 
	
	\section{Challenges in Data Protection}
	Like any other security mechanism, data protection systems also face many challenges when protecting confidential data. There are mainly seven challenges identified for data protection. To implement a successful data protection system, these challenges must be considered and addressed adequately \cite{ICCNSJapan,DPR}. 
	\subsection{Leaking Channels}
	For Data-At-Rest and Data-In-Use, confidential data can be leaked through channels such as USB ports, CD drives, web services, and printed documents. Further, leaking channels associated with Data-In-Motion, such as web services and file sharing, might be extremely challenging, since these channels may be business prerequisites. To ensure maximum security for data passing through these channels, extensive traffic filtering must be carried out.
	\subsection{Human Factor}
	It is always difficult to predict human behavior because it is influenced by many psychological and social factors. Many human actions are affected by subjectivity in making decisions, such as defining the secrecy level of data, assigning access rights to specific users.
	\subsection{Access Right}
	It is important for DLPSs to be able to distinguish between different users based on their privileges and permissions. Without a proper definition of access rights, DLPSs cannot decide whether or not the data is being accessed by a legitimate user, and then it can be misused by a malicious entity.
	\subsection{Encryption and Steganography}
	Network-based DLPSs and DLDSs attempt to identify copies of confidential data using various analysis techniques, and then compare them to the original data. However, the use of strong encryption algorithms makes it very difficult or nearly impossible to analyze the data content. Furthermore, the steganography technique can be used to maliciously leak confidential data since it is highly likely to bypass traffic inspection mechanisms.
	\subsection{Data Modification}
	Some DLPSs and DLDSs use data hashing to check outgoing traffic and to monitor the outgoing data respectively. The problem with hashing is that any modification to the original document can lead to a totally different hash value, resulting in disclosure. Furthermore, some DLDSs use watermarking and steganography to hide the secret data/information that can be modified by some malicious entity.
	\subsection{Scalability and Integration}
	Like many other security mechanisms, data protection systems too can be affected by the amount of data being processed. DLPSs tend to have a poor integration/association with other security mechanisms in a network. This is because some of their mandatory features already exist in other solutions such as firewalls, IDSs, and proxy servers which may arise inconsistencies.
	\subsection{Data Classification}
	If the data is not classified into different levels, data protection systems will not be able to distinguish between confidential and normal data/traffic. Hence, without proper data classification, confidential data can easily be revealed even in the presence of data protection systems.
	
	\section{Data Protection Solutions}
\begin{figure}[!ht]
		\centering
		\includegraphics[width=\textwidth]{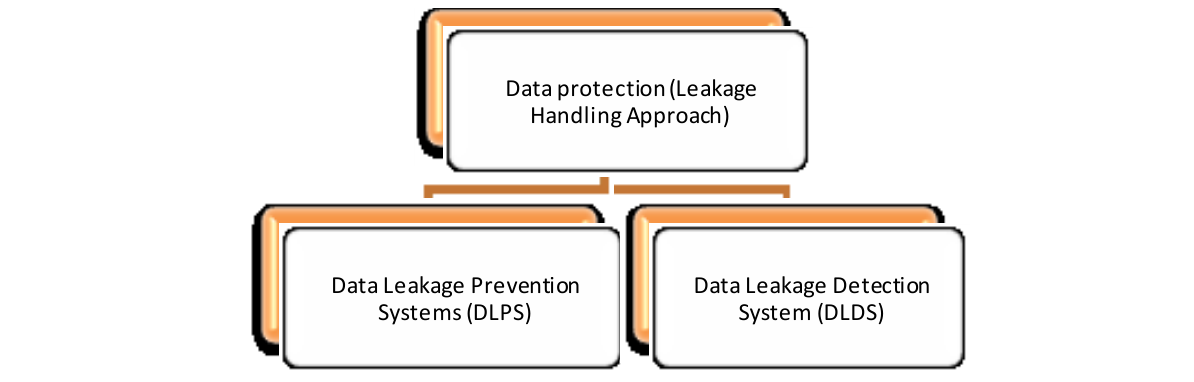}
		\caption{Solution for data protection.}
		\label{fig3}
	\end{figure}

	Data leakage prevention and data leakage detection solution are provided to protect the data as shown in Fig. \ref{fig3}.

	\subsection{Data Leakage Prevention}
	Data leakage/loss prevention (DLP) is typically defined as any solution or process that identifies confidential data, tracks that data as it moves through and out of the enterprise, and prevents unauthorized disclosure of data by creating and enforcing disclosure policies \cite{Confidentiality,Rajat}. 
	
	DLPSs are defined as designated analytical systems used to protect data from unauthorized disclosure at all states using remedial actions triggered by a set of rules. This definition contains three main attributes that distinguish DLPSs from conventional security measures. First, DLPSs have the ability to analyze the content of confidential data and the surrounding context. Second, DLPS scans be deployed to provide protection of confidential data in different states, that is, in transit, in use, and at rest. The third attribute is the ability to protect data through various remedial actions, such as notifying, auditing, blocking, encrypting, and quarantining. The protection normally starts with the ability to detect potential leaks through heuristics, rules, patterns, and fingerprints. The prevention then happens accordingly \cite{MACI,Kaur2017}.
	\subsection{Data Leakage Detection}
	Data leakage detection (DLD) is typically defined as any solution or process that identifies the unauthorized disclosure of confidential data. Data leakage detection systems (DLDSs) are specially designed systems that have the ability to monitor and protect confidential data, detect misuse of confidential data and identify the malicious entity responsible for data leakage \cite{DLD,singh2020survey}.
	
	\section{Data Leakage Protection System}
\begin{figure}[!h]
		\centering
		\includegraphics[width=\textwidth]{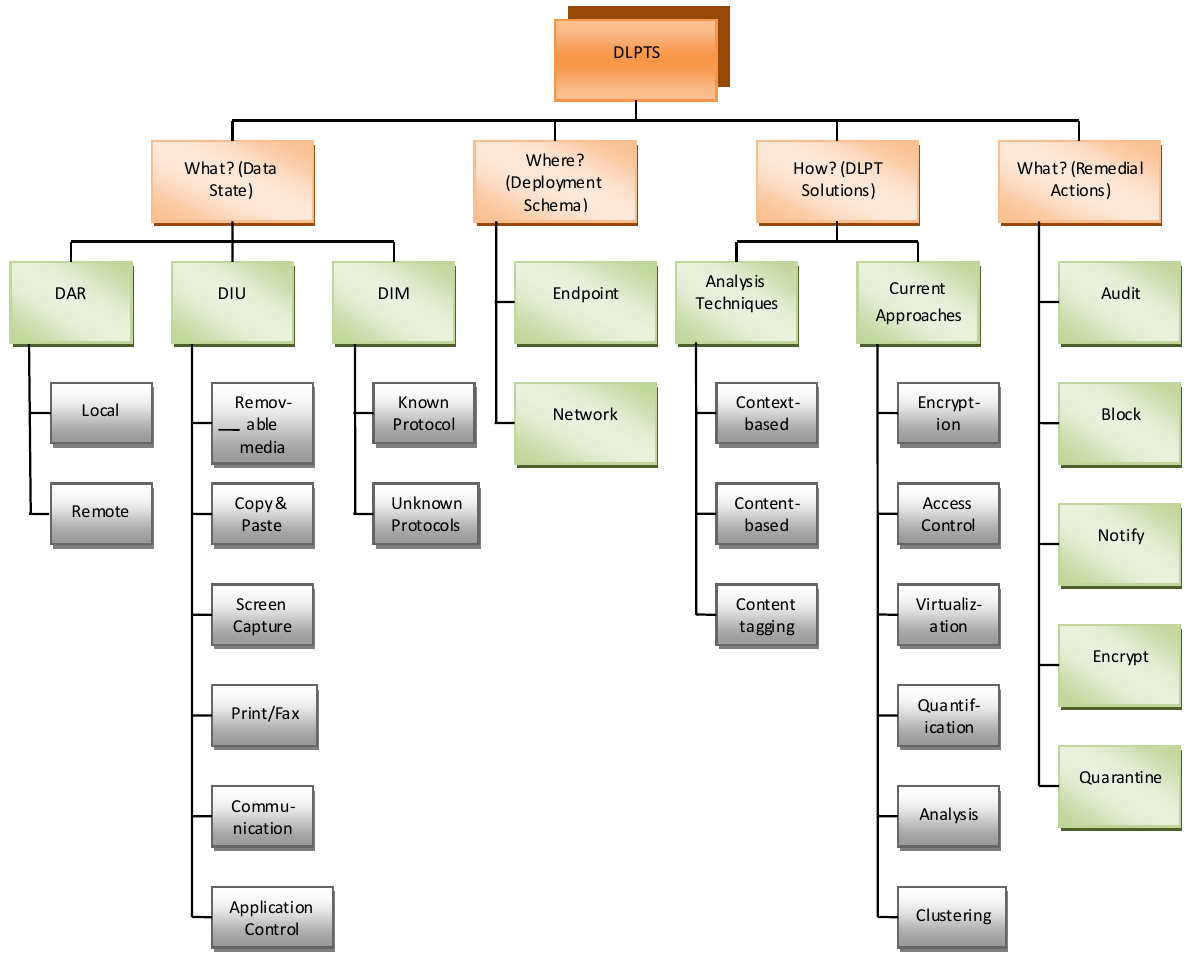}
		\caption{A taxonomy of Data leakage protection system (DLPTS).}
		\label{fig1}
	\end{figure}

	A taxonomy of Data Leakage Protection System (DLPTS) is represented in Fig. \ref{fig1} that represents, what data to be secured, deployment schemes, solution for data protection and the remedial action to be taken. 
	\subsection{Data State}
	Data states include \textquotedblleft Data-At-Rest (DAR)", \textquotedblleft Data-In-Use (DIU)" and \textquotedblleft Data-In-Motion (DIM)". \textit{Data-At-Rest} is the type of data that is stored in repositories either locally or at remote locations. It consists of application databases, backup files, and file systems. It is normally protected by strong access controls, including physical and logical mechanisms. \textit{Data-In-Use} is the data that is accessible to the user in the forms of documents, emails, and applications. This type of data appears in plain text, so it can be easily interpreted and processed. \textit{Data-In-Motion} is the data being transmitted from one node to another. This type of data travels internally between nodes within the same network or externally between nodes that belong to different networks \cite{HybridCCE}. 
	
	\subsection{Deployment Schema}
	DLPS schemes are deployed at the endpoints to protect the data at rest and data in use and in the network to protect the data in motion/transit. Depending on the targeted data for protection, DLPS deployment can take many forms. For example, DLPSs that deal with \textit{Data-At-Rest} are normally focused on protecting known data. The protection comes in the forms of preventing access to data based on predefined security policies. Also, this type of DLPS helps in protecting data at rest by encrypting entire file systems. Furthermore, It scans, identifies, and secures confidential data in data repositories while auditing and reporting security policy violations \cite{Jiangjiang}.
	
	Protecting \textit{Data-In-Use}  requires built-in software that acts like a DLP agent on endpoints. This agent is responsible for disabling or enabling access to applications that deal with confidential data. It is also responsible for blocking confidential data transfer through portable media, that is, CDs, USB drives, and memory cards. Furthermore, it restricts copying, pasting, and editing of confidential data as well as restricting making hard copies through printers. It audits all activities related to confidential data access \cite{Hart}.
	
	For \textit{Data-In-Motion}, DLP appliances are normally used. They come with special processing capabilities to handle large amounts of data. This type of DLPS is responsible for inspecting all outbound traffic for confidential data. It also acts like a proxy server when accessing some applications with confidential data. Moreover, it proactively reports and alerts security administrators and users about potential data leaks. Finally, it coordinates with other security mechanisms such as Secure Sockets Layer (SSL) proxies and network firewalls \cite{Mogull}. 
	
	\subsection{DLPTSs Analysis Technique}
	Context-based DLPTSs focus on the metadata (context), such as size, timing, source, and destination surrounding confidential data, rather than the sensitivity of the content to detect any potential leaks. Content-based analysis DLPSs are more common than and preferable to those that are context-based since it is more logical to focus on the protection of the data itself than on the surrounding context \cite{Mogull}. 
	
	A typical content-based DLPTS works by monitoring sensitive data in its repository or on the go, mainly by using regular expressions, data fingerprinting, and statistical analysis. Regular expressions are normally used under a certain rule such as detecting social security numbers and credit card numbers. The problem with DLPTSs using regular expressions analysis is that they offer limited data protection and have high false-positive rates \cite{Mogull}.
	
	DLPTSs using data fingerprints have better coverage for sensitive data because they have the ability to detect and prevent the leakage of a whole document or parts of a document. However, traditional fingerprinting can lose track when the sensitive data is altered or modified \cite{kaur2017comparative,IOSR}.
	
	Although not widely used in DLPTSs, statistical analysis can work in a fuzzy environment where the sensitive data is not well structured. The main advantage of such a technique is the ability to identify sensitive documents even after extreme modification. In particular, DLPTSs with statistical analysis capabilities can use machine learning algorithms or Bayesian probability to identify altered documents. They can also use text-clustering techniques to construct scattered traces of sensitive data. N-gram analysis and term weighting analysis are the main statistical analysis techniques. N-gram statistical analysis is the way to analyze data based on the frequency of data of interest. It is a method based on machine learning that classifies enterprise documents as sensitive or not \cite{Hart}. Term weighting is a statistical method that indicates the importance of a word within a document. It is normally used in text classification. 
	
	A third technique called content tagging is used in some DLPTSs. This technique is used to tag the file containing confidential data. It can preserve the identity of the file but not the contained confidential data.
	
	\section{Current Approaches of DLPT}
\begin{figure}[!ht]
		\centering
		\includegraphics[width=\textwidth]{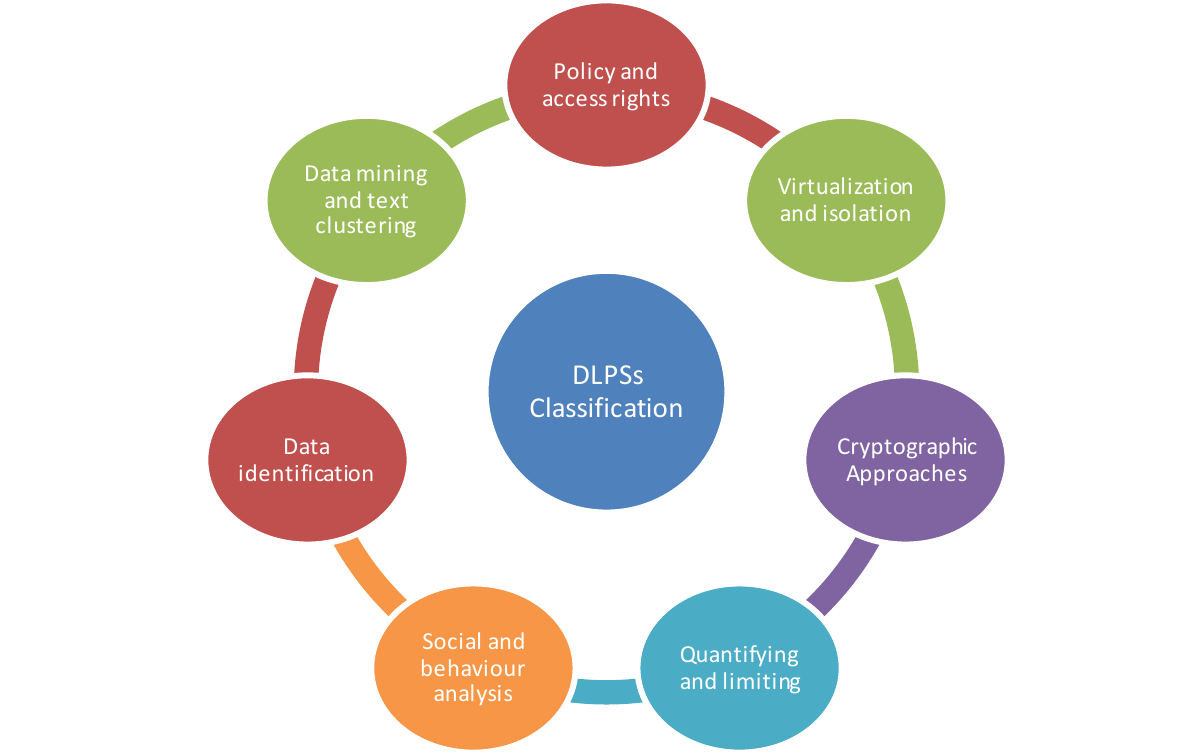}
		\caption{Classification of DLPTSs.}
		\label{fig4}
	\end{figure}

	On the basis of the review, there are mainly seven approaches for data protection as represented in Fig. \ref{fig4} that are summarized in Table \ref{table1} with their strengths and weaknesses.
	
		\begin{table}[!htbp]
			\caption{Strength and weakness of current DLPT approaches.}
			\label{table1}
			\centering 
			\begin{tabular}{p{.15\textwidth}p{.38\textwidth}p{.38\textwidth}}
				\hline 
				\textbf{Approach} & \textbf{Strengths} & \textbf{Weaknesses} \\ 
				\hline \vspace{0em}
				Policy and access rights&\vspace{-0.5em}\begin{itemize}[leftmargin=*]
					\item suitable for any organization if access rights and data classification are properly established 
					\item easy to manage 
					\item suitable for Data-At-Rest and Data-In-Use 
					\item strong prevention mechanism 
				\end{itemize}
				& \vspace{-0.5em}
				\begin{itemize}[leftmargin=*]
					\item affected by improper data classification 
					\item affected by the access control policy in use 
					\item not a detective method, hence if a leak is happening the method is ineffective
				\end{itemize}	   \\ 
				\hline \vspace{0em}
				Virtualization and isolation&\vspace{-0.5em}\begin{itemize}[leftmargin=*]
					\item requires small hardware implementation 
					\item dynamic as it does not need regular administrative interference 
					\item accessing sensitive data can use existing data classification
				\end{itemize}  & \vspace{-0.5em} \begin{itemize}[leftmargin=*]
					\item not mature enough
					\item produces considerable amount of overheads
					\item not a detection method
				\end{itemize}\\ 
				\hline \vspace{0em}
				Cryptographic approaches& \vspace{-0.5em} \begin{itemize}[leftmargin=*]
					\item strong cryptography can produce maximum security
					\item cryptographic methods are wieldy to use and have many options
				\end{itemize} & \vspace{-0.5em}
				\begin{itemize}[leftmargin=*]
					\item cryptography can secure sensitive data but may not deny its existence
					\item does not detect data leak
					\item confidential data can be accessed by weak credentials
				\end{itemize} \\ 
				\hline \vspace{0em}
				Quantifying and limiting& \vspace{-0.5em} \begin{itemize}[leftmargin=*]
					\item goes beyond studying sensitive data, and focus on the leaking channels
					\item useful against specific types of attacks such as salami attacks
					\item effective for all data states
				\end{itemize} & \vspace{-0.5em}\begin{itemize}[leftmargin=*]
					\item does not ensure total blockage of the leaking channel
					\item limited to specific situations or scenarios
					\item can disrupt workflow
				\end{itemize} \\ 
				\hline \vspace{0em}
				Social and behavior analysis& \vspace{-0.5em} \begin{itemize}[leftmargin=*]
					\item proactive data leakage prevention by detecting malicious relations
					\item suitable for all data states
				\end{itemize} & \vspace{-0.5em} \begin{itemize}[leftmargin=*]
					\item produces a high level of false-positive requires regular administrative interference
					\item requires a huge amount of profiling and indexing
				\end{itemize} \\ 
				\hline \vspace{0em}
				Data identification& \vspace{-0.5em} \begin{itemize}[leftmargin=*]
					\item very strong in detecting unmodified data
					\item very low false positive level for analysis using fingerprints
					\item some robust hashing can detect modified data
				\end{itemize}   & \vspace{-0.5em} \begin{itemize}[leftmargin=*]
					\item extremely modified data cannot be detected
					\item lacking semantic understanding
				\end{itemize} \\ 
				\hline \vspace{0em}
				Data mining and text clustering& \vspace{-0.5em} \begin{itemize}[leftmargin=*]
					\item can predict future data leaks powerful in detecting unstructured data
					\item less dependent on administrative help flexible and adaptable
				\end{itemize}  & \vspace{-0.5em} \begin{itemize}[leftmargin=*]
					\item requires a great deal of processing requires learning phase, which means many false positives
				\end{itemize} \\ 
				\hline 
			\end{tabular} 
	\end{table}
	
	The models that handle with data leakage problem by embedding the code in the document are presented in \cite{kumarneeraj,Backes,Kamal}. Backes et al. provide a generic data lineage framework called LIME for data flow across multiple entities having two principal roles i.e. owner and consumer. The guilty entity is identified by developing a data transfer protocol between two entities via considering an oblivious transfer, watermarking, and signature primitives. This method considers the possible data leakage and the corresponding constraints at the design stage. A model that identifies the guilty party who is responsible for leaking critical organizational data is given by Kumar et al. This method is able to find the exact client who leaked the data but it is unable to detect the guilty client if embedded code is modified or completely destroyed by the client.
	
	The model introduced in \cite{Papadimitriou,shabtai2012survey} helps to identify the guilty agent who leaks the data provided by the distributor. In \cite{Papadimitriou} the model assesses the guilt of various agents. It consists of various algorithms for the distribution of data among multiple agents which improve the chances of identifying a leaker and then the fake objects are added to the distributed set. Data is allocated to various agents in such a manner that there will be minimum overlapping of data among various agents. Later on, if any agent leaked his confidential data at some unauthorized place then the guilty party can be found on the basis of the data allocated to the agents. Kumar et al. introduce the allocation strategies that work on the basis of No- wait model and increase the chances of identifying the guilty party. These methods have the benefit that the agent can not destroy the information after leaking the data as in the case of watermarking but it cannot identify the exact data leaker in the case when the same data is allocated to multiple agents or overlapping of data among agents increases. 
	
	The method introduced in \cite{LiuFang195} enables the data owner to depute the detection operation safely to a semi-host provider without exposing the critical data to the provider. This method is implemented using the fuzzy fingerprint technique to elevate data privacy at the time of data leak detection operations. The method provides the privacy-preserving Data Leakage Detection (DLD) solution to handle data leaks in situations where a set of sensitive data digests is utilized for detection.
	
	To protect sensitive information from unauthorized parties, Bishop et al. present an information leakage detection (ILD) agent system in \cite{Bishop}. It is a mobile agent-based approach that brutalizes the process of discovering and coloring perceptive hosts file systems and observing the colored file system for detection of potential information leakage. A concept presented is based on assigning a sensitivity score to data sets for mitigating the data misuse and data leakage incidents in database system \cite{Harel}.
	
	Other solution for data leaks includes sequence alignment techniques \cite{ShuPrivacy-Preserving,ShuFastDetection,ShuRapid, ShuRP}, classification technique \cite{Guevara}, network-based data leak detection solution \cite{ShuSecureComm,Sodagudi,EPS}. In \cite{ShuFastDetection} solution for the detection of transformed data leaks is presented by detecting complex data leak patterns using sequence alignment techniques. A new class of health information security intrusions that exploits physiological information leakage to compromise privacy is presented in \cite{Nia}. A map-reduce framework is introduced by Liu et al. for the detection of exposed sensitive contents in \cite{LiuFang195}. A network-wide method based on black-box differencing for restraining and controlling the flow of confidential data within a network is presented in \cite{Croft,Jung}, which determines when the secret data is being leaked. \cite{AlneyadiJNCA,Hauer,AlneyadiDLPA} present the way for data leakage prevention. In \cite{shabtai2012survey},  it is specified that solutions for data leakage preventions are mainly deployed to prevent unintentional leakage incidents. Table \ref{table2} summarizes the most relevant work of the field.
	
		\begin{longtable}{|p{.15\textwidth}|p{.12\textwidth}|p{.10\textwidth}|p{.10\textwidth}|p{.18\textwidth}|p{.18\textwidth}|}
			\caption{Summary of most relevant work in the review.}\\
			\hline 
			\textbf{	Paper and Category}& \textbf{Method} & \textbf{Analysis} & \textbf{Suitable for} & \textbf{Contribution} & \textbf{Limitation} \\ 
			\hline 
			Wuchner et al. \cite{Wuchner}, 2012 (Policy and access rights)& Detective/ Preventive & Context & In use &UC4Win, a data loss prevention solution for Microsoft Windows operating systems.  &Requires predefined policy. Cannot identify sensitive data.  \\ 
			\hline 
			Squicciarini et al. \cite{Squicciarini}, 2010 (Policy and access rights)	& Preventive & Context & In use &Introduces a three layers data protection framework.  &Requires a pre-defined classification for data. Mis- classified sensitive data can be leaked.  \\ 
			\hline 
			Griffin et al. \cite{Griffin}, 2005 (Virtualization and isolation)&Preventive  & Context & In use & Proposes Virtual Trusted Domains (VTD) to offload processes to secure environments. & Imposes challenge to computational capabilities. \\ 
			\hline 
			Wu et al. \cite{Jiangjiang}, 2011 (Virtualization and isolation/ Cryptographic approaches)&Preventive  & Context & At rest/ In use &Introduces a combination of encrypted storage and virtual environment to prevent data leakage.  & Suitable for data at rest only. Cannot prevent data leakage caused by privileged used. \\ 
			\hline 
			Blanke \cite{Blanke}, 2011 (Cryptographic approaches)& Preventive & Context & At rest & Uses ephemeral encryption to protect data whenever accessed by a user. & Requires a pre-implementation of an encrypted file system. \\ 
			\hline 
			Yoshihama et al. \cite{Yoshihama}, 2010  (Quantifying and limiting)& Detective & Content/ Context & In motion & Uses an application-level proxy to detect potential data leakage risks. &Cannot detect data leaked through covert channels.  \\ 
			\hline 
			Borders et al. \cite{Borders}, 2009 (Quantifying and limiting)& Detective & Content & In motion & Constrains the maximum volume of sensitive data in web traffic. & Unable to filter parts of URLs that contains random numbers to prevent cashing. \\ 
			\hline 
			Suen et al. \cite{Suen}, 2013 (Quantifying and limiting/ Social and behavior analysis)	& Detective  & Context & In motion &Uses S2 Logger to track files while traveling in the cloud.  & Based on content tagging. Cannot track sensitive content. \\ 
			\hline 
			Boehmer \cite{Boehmer}, 2010 (Social and behavior analysis)	& Detective/ Preventive &Context  & In use & Uses case-based reasoning (CBR) in combination with directed acyclic graph (DAG) and Hamming similarity function. & Needs existing or synthetic compliance profiles for comparison process. \\ 
			\hline 
			Shapira et al. \cite{Shapira}, 2013 (Data identification)	& Detective & Content & In use & Robust fingerprinting to overcome shortcoming in ordinary hashing. & Requires extensive data indexing for both sensitive and normal data. \\ 
			\hline 
			Shu et al. \cite{ShuSecureComm}, 2013 (Data identification)	& Detective & Content &In motion  &Uses message shingles/fuzzy fingerprints to detect in advertent data leak in network traffic.  & Modified data can cause false negatives because the shingles fingerprints are different from the original ones. \\ 
			\hline 
			Hart et al. \cite{Hart}, 2011 (Data mining and Text clustering)	& Detective & Content & In use & Uses SVM Machine learning to classify documents to private and public. & Inflexible, Limited to two categories. \\ 
			\hline 
			Lindell et al. \cite{Lindell}, 2000 (Data mining and Text clustering)	& Preventive &Content  & At rest/ In use & Sharing confidential data on the union of the entities databases, without releasing unnecessary information. & Theoretically proven but lacking practical experiments. \\ 
			\hline 
			Marecki et al. \cite{Marecki}, 2010 (Data mining and Text clustering)	& Detective &Context  & In motion & Uses Partially Observed Markov Decision Processes (POMDPs) over decision epochs. & POMDP requires huge amount of calculations. \\ 
			\hline 
			Gomez-Hidalgo et al. \cite{Hidalgo}, 2010 (Data mining and Text clustering)	&Detective/ Preventive  &Content  &In motion  & NER (named entity recognition) approach is used to identify and extract words from texts. &  Named entity recognition could be affected by spelling mistakes and connected words.\\ 
			\hline 
			Sokolova et al. \cite{Sokolova}, 2009 (Data mining and Text clustering)	& Detective &Content  & In motion & Uses support vector machine to classify enterprise documents as sensitive non-sensitive.  & Not fixable because it classifies data to public or private only.  \\ 
			\hline 
			Parekh et al. \cite{Parekh}, 2006 (Data mining and Text clustering)	& Detective & Content  & In motion & A new approach to enable the sharing information of suspicious payloads. &  Polymorphic/ obfuscated worms and mimicry attacks may create a big challenge.\\ 
			\hline 
			Carvalho et al. \cite{Carvalho}, 2009 (Data mining and Text clustering)	&Detective  & Content/ Context &In motion  &Presents an extension –Cut Once- to \textquotedblleft Mozilla Thunderbird".  & Introduces high level of false positives, since it requires existing messages in the sent folder. \\ 
			\hline 
			Zilberman et al. \cite{Zilberman}, 2010 (Data mining and Text clustering/ Social and behavior analysis)	& Detective/ Preventive & Content/ Context & In motion & Uses TF-IDF and cosine similarity to compute existing links between users. & High false positive rates because of short history between senders and recipients. \\ 
			\hline 
			Carvalho and Cohen \cite{CarvalhoCohen}, 2007 (Data mining \& Text clustering/ Social \& behavior analysis)	& Detective &  Content/ Context& In motion &Predicts unintended message recipients.  & Can create false outliers because of limited interaction history.\\\hline 
			\label{table2}
	\end{longtable}
	
	\section{Emerging Challenges and Future Research Directions}
	Data protection mechanism should make sure that the data is safe enough from all the internal as well as external threats so that there should not be any problem such as loss of data or data theft. Many methods and solutions have been proposed to protect the data from leakage/loss but still there exist some research gaps in the provided solutions. We need an ideal solution that can secure, manage confidential data and locate the malicious agent and activities. The following research gaps are identified based on the literature review:
	
	\begin{enumerate}
		\item \textit{Increased Overhead:} When the data is stored at a third party or somewhere else in the database or repositories, the security mechanism such as encryption scheme, etc. are applied to the whole data which reduces the utility of data and increase the overhead of the service provider, Furthermore some of the approaches require a huge amount of calculations resulting in delay and increased overhead.
		\item \textit{Static Request handling:} Solution has been provided for the detection of a malicious agent by providing a guilty agent model. 
		\item \textit{Single Objective Approaches:} The approaches provide the solution either for detection of guilty agent or prevention of loss of data. These approaches are suitable for data-at-rest or data-in-use or data-in-motion. Some approaches provide prevention solutions but cannot identify data as sensitive. For the better protection of data, there is a need for a hybrid approach that fulfills all the objectives.
		\item \textit{Involves Data modification:} The guilty agent is identified by hiding the information in the requested document which may involve modification of data. Furthermore, DLP uses techniques such as regular expression, fingerprinting, data mining/ text clustering, etc. to identify the sensitive data, but if the information is too much modified then these techniques cannot extract the sensitive data.
		\item \textit{High false-positive rate:} The approaches used for DLP require a learning phase that may involve a high false-positive rate. 
	\end{enumerate}
	
	The following are the research directions in order to fill the identified research challenges:
	\begin{enumerate}
		\item To make a balance between data utility and security and reducing overheads by proper classification.
		\item To formulate a Data leakage Detection System that handles the user request in a dynamic manner. 
		\item Efforts will be made in developing the system with high accuracy.
		\item Development of a new multi-objective approach that provides both Data Leakage Detection and Prevention
	\end{enumerate}
	
	\section{Conclusions and Future Scope}
	Prevention of data disclosure to unauthorized entities is one of the main goals of information security. According to datalossdb ($2015$) report, in the year $2014$, about $50\%$ of recorded data leakage occurred in the business sector, about $20\%$ occurred in the government sector and about $30\%$ occurred in the health and education sectors. Although some reported leaks were not detrimental to organizations, others have caused several million dollars worth of damage. Business credibility is compromised when sensitive data is leaked to competitors. The developments to be carried out in research work will be helpful in business, governmental, private, etc. sectors with the following advantages:	
	\begin{inparaenum}[(i)]
		\item Provide Data protection.
		\item Maintains Privacy and Security of data.
		\item Mitigates the risk of data leakage.
		\item Solves the controversy raised due to information leakage.
	\end{inparaenum}
	
	%\bibliographystyle{IEEEtran}
	%\bibliography{mybibfile}

\end{document}